\documentclass[fleqn,usenatbib]{mnras}

\usepackage{newtxtext,newtxmath}
\usepackage[T1]{fontenc}

\DeclareRobustCommand{\VAN}[3]{#2}
\let\VANthebibliography\thebibliography
\def\thebibliography{\DeclareRobustCommand{\VAN}[3]{##3}\VANthebibliography}

\usepackage{graphicx}	% Including figure files
\usepackage{amsmath}	% Advanced maths commands

\usepackage{amssymb}	% Extra maths symbols
\usepackage{caption}
\title[Lensed LAEs in eBOSS]{LESSER: A Catalogue of Spectroscopically selected sample of Lyman-$\alpha$ Emitters Lensed By Galaxies}

\author[Xiaoyue Cao et al.]{
Xiaoyue Cao,$^{1,2}$
Ran Li$^{2,1}$\thanks{E-mail: ranl@bao.ac.cn (NAOC)},
Yiping Shu,$^{3}$
Shude Mao,$^{4,2}$
Jean-Paul Kneib,$^{5,6}$
Liang Gao$^{2,1}$
\\
$^{1}$School of Astronomy and Space Science, University of Chinese Academy of Sciences, Beijing 100049, China\\
$^{2}$National Astronomical Observatories, Chinese Academy of Sciences, 20A Datun Road, Chaoyang District, Beijing 100012, China\\
$^{3}$Institute of Astronomy, University of Cambridge, Madingley Rd, Cambridge, CB3 0HA, UK\\
$^{4}$Department of Astronomy, Tsinghua University, Beijing 100084, China\\
$^{5}$Institute of Physics, Laboratory of Astrophysics, Ecole Polytechnique Fdrale de Lausanne (EPFL), Observatoire de Sauverny, 1290
Versoix, Switzerland\\
$^{6}$Aix Marseille Université, CNRS, LAM (Laboratoire d’Astrophysique de Marseille) UMR 7326, 13388, Marseille, France
}

\pubyear{2020}

\begin{document}
\label{firstpage}
\pagerange{\pageref{firstpage}--\pageref{lastpage}}
\maketitle
\begin{abstract}
We introduce the LEnSed laeS in the Eboss suRvey (LESSER) project, which aims
to search for lensed Lyman-$\alpha$ Emitters (LAEs) in the Extended Baryon Oscillation Spectroscopic Survey (eBOSS). The final catalog contains 361 candidate lensing systems. The lens galaxies are luminous red galaxies (LRGs) at redshift $0.4 < z < 0.8$, and the source galaxies are LAEs at redshift $2 < z < 3$. The spectral resolution of eBOSS ($\sim$2000) allows us to further identify the fine structures of Lyman-$\alpha$ ($\rm Ly\alpha$) emissions. Among our lensed LAE candidates, 281 systems present single-peaked line profiles while 80 systems show double-peaked features. Future spectroscopic/imaging follow-up observations of the catalog may shed light on the origin of diverse $\rm Ly\alpha$ line morphology, and provide promising labs for studying low mass dark matter haloes/subhaloes. 
\end{abstract}
%LEnSed laeS in Eboss suRvey(LESSER)


%%%%%%%%%%%%%%%%%%%%%%%%%%%%%%%%%%%%%%%%%%%%%%%%%%

%%%%%%%%%%%%%%%%% BODY OF PAPER %%%%%%%%%%%%%%%%%%

\section{Introduction}
The so-called ``spectroscopic method'' is a very effective way to search for galaxy-scale strong lensing systems \citep{Warren1996,Willis2005,Willis2006,Bolton2004}. The basic idea is to scan the spectra of foreground galaxies for abnormal emission lines that actually come from background source galaxies. Detection of such ``abnormal'' emission lines provides an evidence that two galaxies are nearly aligned along the line of sight, thus are likely a strong lensing system. Based on the spectroscopic method, the Sloan Lens ACS Survey \citep[SLACS,][]{Bolton2008}, the Baryon Oscillation Spectroscopic Survey (BOSS) Emission-Line Lens Survey \citep[BELLS,][]{Brownstein2012}, and the SLACS for the Masses Survey \citep[S4TM,][]{Shu2015,Shu2017} have achieved many successes, extending the number of galaxy-galaxy strong lensing systems to $\sim$200, and providing valuable constraints on the galaxy formation theory and cosmological models \citep{Treu2006,Koopmans2006,Koopmans2009,Gavazzi2007,Jiang2007,Bolton2012, chenyun2019}. Recently, \citet{xinlun2020} finds another $\sim$100 possible lensed $\rm [OII]$ emitters through the spectroscopic method by searching the full BOSS database.

\citet{Shu2016a} initiated the BELLS for the GALaxy-Ly$\alpha$ EmitteR sYstems Survey (the BELLS GALLERY Survey, hereafter) project, which aims at detecting galaxy-scale gravitational lenses with high-redshift LAEs as the sources. They find 187 possible galaxy-LAE systems from millions of spectra in the BOSS database \citep{Alam2015}, 21 most promising candidates among them have follow-up imaging observations with the Hubble Space Telescope (HST), and 17 candidates are confirmed as real lensing systems \citep{Shu2016b}. With the help of lensing magnification, these lensed LAEs provide information on the size and morphology of sub-kiloparsec scales clumps in LAEs \citep{Shu2016b,Ritondale2019b,Cornachione2018}, and a promising lab for constraining the identity of dark matter \citep{ranli2016, ranli2017, Ritondale2019a}. 

Encouraged by the success of the BELLS GALLERY survey, we apply the similar searching technique to the Extended Baryon Oscillation Spectroscopic Survey \citep[eBOSS,][]{Dawson2016}. The eBOSS survey is part of the Sloan Digital Sky Survey-IV (SDSS-IV) project \citep{Blanton2017}, which started in 2014 using the 2.5-meter Sloan telescope \citep{Gunn2006} at the Apache Point Observatory in New Mexico. The main difference between the method in this work and that in \citet{Shu2016a} is the treatment to the $\rm Ly\alpha$-emission profile. \citet{Shu2016a} didn't classify the line profile of lensed LAEs,  although the spectroscopic resolution of BOSS ($\sim$2000) provides such ability. In this work, we explicitly classify our lensed LAE candidates into single-peaked and double-peaked types.

%\xycao{The main difference between the method in this work and that in \citet{Shu2016a} is the treatment to the $\rm Ly\alpha$-emission profile. \citet{Shu2016a} don't classify the line profile of lensed LAEs,  although the spectroscopic resolution of BOSS provides such ability. }

%We visually inspect the spectra of samples in \citet{Shu2016a}, while the final samples for HST Follow-up are all single-peaked LAEs, some candidates in their parent catalog ($\sim$187 candidates), although not clearly announced, show the double-peaked $\rm Ly\alpha$ emission line.} In this project, we explicitly classify our lensed LAE candidates into single-peaked and double-peaked types.

%In \citet{Shu2016a}, the parent candidate catalog does not explicitly distinguish between single-peaked and double-peaked line profiles and the final candidates with HST follow-up imaging are all single-peaked LAEs. The spectral resolution of eBOSS ($\sim$2000) allows us to further identify the fine structures of Lyman-$\alpha$ ($\rm Ly\alpha$) emissions \citep{Smee2013}. In this project, we explicitly classify our lensed LAE candidates into single-peaked and double-peaked types.

Previous spectroscopic observations have shown that those double-peaked LAEs are quite common at z$\sim$3 \citep{Yamada2012,Tapken2007}. Theoretically, internal gas dynamics of the LAE with double-peaked $\rm Ly\alpha$ emission is different from the single-peaked one \citep{Yanghuan2017, Gronke2016,Gronke2015,Orsi2012,Dijkstra2006,Ahn2004,Neufeld1990}. Therefore a complete sample that includes both types of lines can provide a more comprehensive picture of the LAEs.

Hereafter, we refer to this project as ``LESSER'' (LEnSed laeS in the Eboss suRvey). This paper is organized as follows. We introduce the spectroscopic identification pipeline for lensed LAEs in Section 2. The catalog of candidates and their statistical properties are shown in Section 3. Finally, in Section 4 we discuss prospects for future applications of our sample.

\section{The searching pipeline}
The spectroscopic method of identifying lens systems has been well described in previous works \citep{Bolton2004,Brownstein2012}. \citet{Shu2016a} extended this method to search for lensed LAEs. In this paper, we apply the same technique as that used in \citet{Shu2016a} to the eBOSS data \citep{Ahumada2019}, with modifications to distinguish between single-peaked and double-peaked Ly$\alpha$ emission lines. We briefly describe the pipeline below.

We start with selecting the galaxies tagged as `Luminous Red Galaxy' by the eBOSS pipeline \citep[LRG,][]{Prakash2016}. The spectra of LRGs are optimal places to search for abnormal emissions from background sources for two reasons: First, LRGs are massive early-type elliptical galaxies, thus can act as efficient lenses; Second, the spectra of LRGs typically contain few emission lines, and their continuum can be well modeled by the template fitting \citep{Blanton2017,Bolton2012b}. There are 1571129 LRG spectra selected in this step.

For each LRG spectrum, we mask the regions that correspond to the empirically determined emission, absorption, and sky-line features\footnote{We list the wavelength of all those line-region in a Table that can be found on \url{https://github.com/caoxiaoyue/search_lensed_laes/blob/master/paper_material/line_info.txt}}. We use the galaxy template generated by eBOSS pipeline \citep{Blanton2017,Bolton2012b} to fit the masked spectrum, then we subtract the best-fit LRG spectra from the original LRG spectrum. The residual reflects the `features' that can not be captured by the galaxy-template of LRG, which may contain `abnormal' emission features from another background object.

\begin{figure}
\includegraphics[width=\columnwidth]{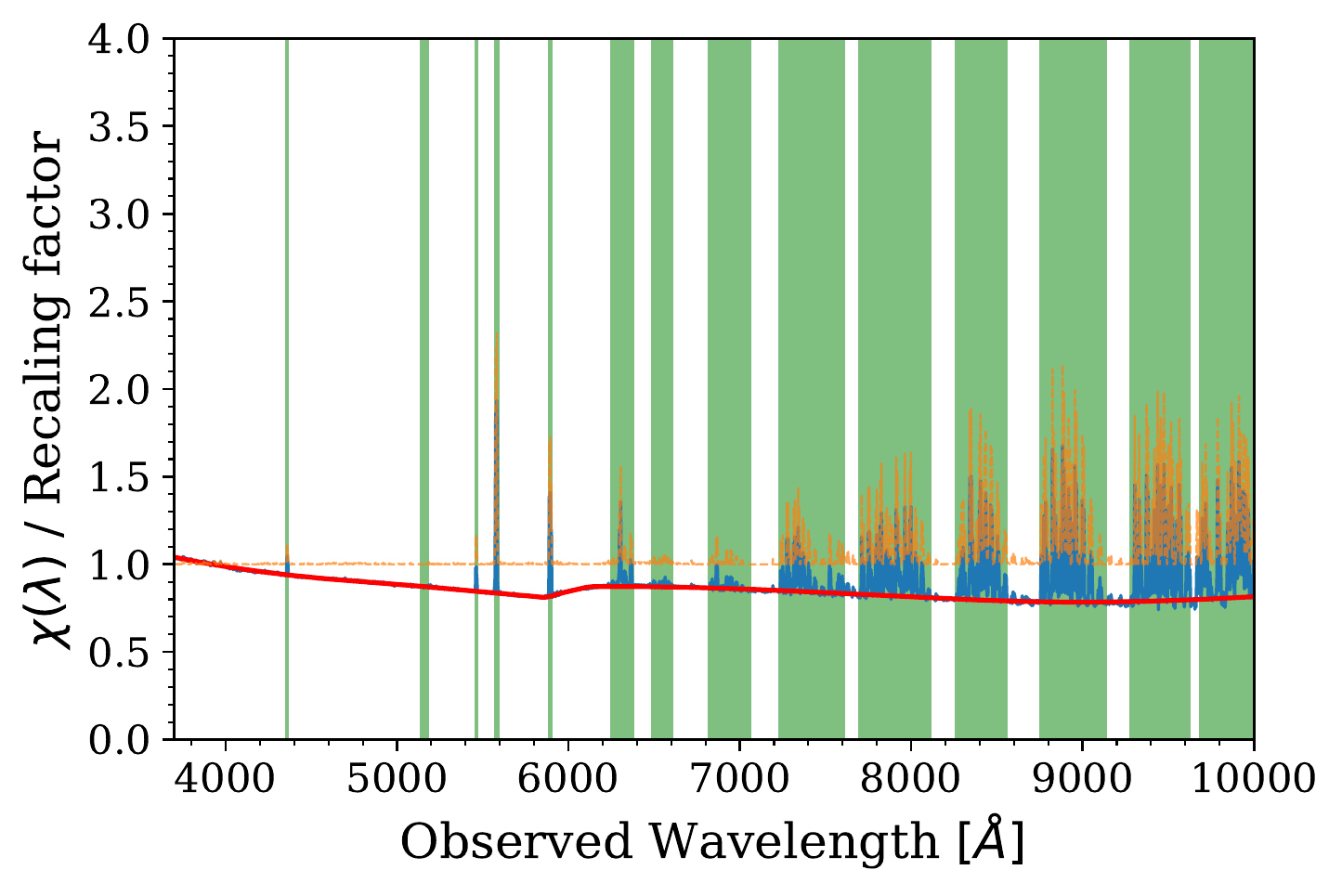}
\caption{The blue line shows the root mean square (RMS) spectrum of the sky-subtracted sky-fibers, which empirically indicate the sky-noise level. If the eBOSS pipeline estimates the sky-noise level perfectly, the blue line would be a horizontal line with an ordinate of $\sim$1. The solid blue line is noisy around the green shadow-region because of the influence of sky-emissions. The red line represents the baseline level of the blue line, which is obtained from B-spline fitting. Divide the blue line by red line; we can estimate how much sky-noise is underestimated, as shown by the orange line.}
\label{fig:noise_rescaling}
\end{figure}

To detect the emission from background source galaxies, we apply an error-weighted match filter on the residual spectrum with a Gaussian kernel, the velocity dispersion of which is set to 150 km/s. We confine our scanning wavelength range to $[3700-4800]$ \AA \ (roughly at redshift $z \simeq 2-3$ for $\rm Ly\alpha$ emission), so that the only prominent foreground contamination emission lines is the $\rm[OII]$ 3727 doublet. We retain the detected emissions (`hit')  with S/N (Signal to Noise ratio) greater than 6. Note that, The skylight subtraction procedure in the eBOSS official pipeline underestimates the noise level at the wavelength location with strong air-glow emission. We correct this effect empirically using sky-subtracted sky-fiber spectra \citep{Brownstein2012,Shu2016a}. Our noise re-calibration result is shown in Figure \ref{fig:noise_rescaling}. There are 4995 hits detected in this step.

We bin the `hits' by observed wavelength. The overpopulated bins are likely to be related to the sky-line signal, thus hits in these bins are pruned. Similarly, we also prune the hits that are overpopulated at rest-frame wavelength, which are associated with imperfect galaxy template fitting. 2813 hits are removed in this step.

\begin{figure}
\includegraphics[width=\columnwidth]{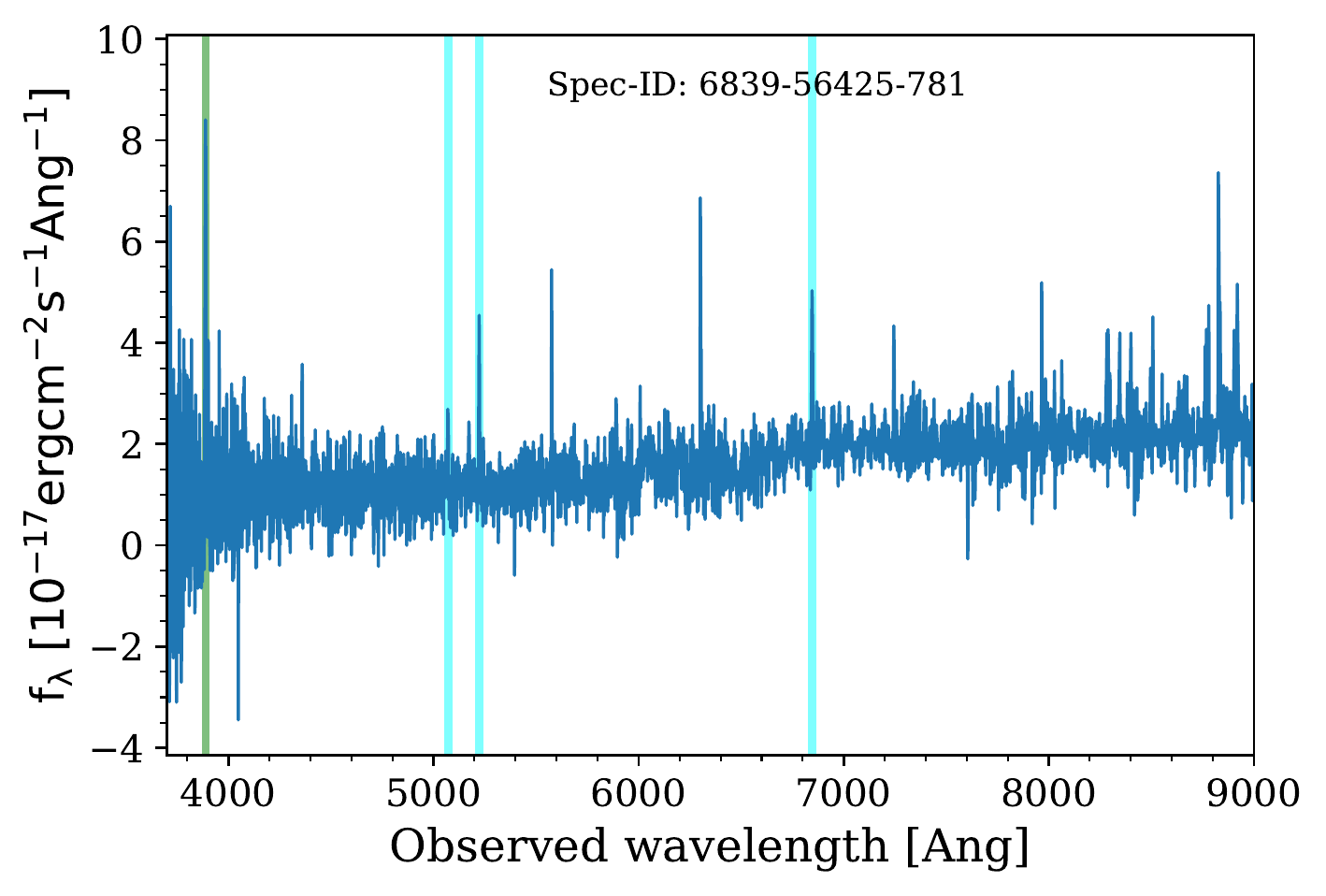}
\caption{This example shows the case where an $\rm [OII]$ emission is misidentified as $\rm Ly\alpha$. The shaded green marks the location of hit, while shaded cyan regions correspond to the position of H$\alpha$, $\rm [OIII]$ 5007, and $\rm H\beta$ emission assuming hit is $\rm [OII]$ emission instead $\rm Ly\alpha$. The legend represents SDSS plate-mjd-fiber.}
\label{fig:oii}
\end{figure}

The $\rm [OII]$ 3727 emission from the low-redshift emitters may appear in our searching range (3700\AA-4800\AA) as a contaminator. From the observational point of view, if the $\rm [OII]$ 3727 emission is shown, the $\rm H\alpha$, $\rm [OIII]$ 5007, and $\rm H\beta$ emissions will likely appear simultaneously. Therefore, for each hit,  we inspect secondary emission lines at wavelength corresponding to H$\alpha$, $\rm [OIII]$ 5007, and $\rm H\beta$, assuming that the hit is $\rm [OII]$ 3727. If the quadrature sum of S/N values at these three wavelength locations is higher than 2.5$\sigma$, the hit is identified as $\rm [OII]$ emission and rejected. An example of the rejected hits is shown in Figure \ref{fig:oii}. There are 921 hits removed by this criterion.

Some `fake hits' may still exist in our candidates. We list the source of the spurious emission and our visually removal strategy as follows:
\begin{enumerate}
\item{Visually inspect the results of galaxy template fitting, remove the spurious emission detection due to bad fit.}
\item{Check the spectra of `neighbor fibers,' which are geometrically near to the fiber with `hit' detection (`hit-fiber') in the SDSS plate. If there is a stronger emission appearing in `neighbor fiber' and its wavelength is the same as the `hit-fiber.' Then the hit is considered as a `false-positive' detection induced by neighboring fibers. Technically, this is called `ccd cross-talk'.}
 \item{The outlier induced by cosmic rays can be identified by checking each sub-exposure spectrum (15 minutes per exposure). The emission that does not appear in all exposures is due to cosmic rays.}
 \item{Visually checking the CCD raw data to ensure the detection is not from CCD defects.}
 \item{Remove the hits that have a line width less than the spectral resolution of the instrument ($\sim70 km/s$).}
 \item{Remove repetitive candidates that have been independently observed multiple times.}
\end{enumerate}
%We have 639 hits left after these cuts.
We have 527 hits left after these cuts.

In order to classify our candidates into double and single-peaked types, We first apply a Gaussian match filter with a smaller dispersion of 75 km/s to the spectrum to reduce the noise; then we search the blueward side of the main-peak for another sub-peak with S/N>2.5, candidates with another blueward bump are tagged as "double-peaked LAE".

Previous observation has revealed the $\rm Ly\alpha$ emission of LAEs typically shows positive skewness \citep{Fuller2020,Vivian2015,Yamada2012}. The asymmetric Ly$\alpha$ line profile of LAEs can be well described by the skew-normal function with the formula form in \citet{Vivian2015} \citep[see also,][]{Mallery2012},
\begin{equation}
\begin{split}
f(\lambda) = A e^{-0.5\times((\lambda-x)/\omega)^2}  [\int_{-\infty}^{s(\lambda-x)/\omega}\exp(-t^2/2)dt] + c \,,
\end{split}
\label{eq0}
\end{equation}
where $s$ is the skewness. The line centre is given by $\lambda_0 = x +\omega\delta\sqrt{2/\pi}$ and the line width is given by $\sigma = \omega\sqrt{1 - 2\delta^2/\pi}$, $\delta = s / \sqrt{1+s^2}$. \citet{Vivian2015} systematically investigate the morphology of $\rm Ly\alpha$ line profile, based on 304 LAE samples at 3<z<7 in the Great Observatories Origins Deep Survey and Cosmic Evolution Survey field. They find the $\rm Ly\alpha$ emission of LAEs has a median skewness value of $\sim$1.5. To increase the purity of our final catalog, we apply a cut on the shape of line profile, removing candidates whose main-peak profiles have negative skewness. We have 361 candidates left in our final catalog.
%\citet{Vivian2015} systematically investigate the morphology of $\rm Ly\alpha$ line profile, based on 304 LAE samples at 3<z<7 in the Great Observatories Origins Deep Survey and Cosmic Evolution Survey field. They find the $\rm Ly\alpha$ emission of LAEs has a median skewness value of $\sim$1.5. To increase the purity of our final catalog, we apply a cut on the shape of line profile, removing candidates whose main-peak profiles have negative skewness. We have 361 candidates left in our final catalog.

\section{Candidate catalog and results}
\label{sec:res}

In total, we find 361 lensed LAE candidates from the eBOSS database, 281 candidates of which are single-peaked, and 80 candidates are double-peaked. The detailed information of our candidates is listed in Tables \ref{tab:single_tab} and \ref{tab:double_tab}. Since the eBOSS database also includes previous BOSS data, we mark candidates that were already found by the BELLS GALLERY project using star symbols in the last column of Tables \ref{tab:single_tab} and  \ref{tab:double_tab}. 

\begin{figure*}
	\includegraphics[width=\textwidth]{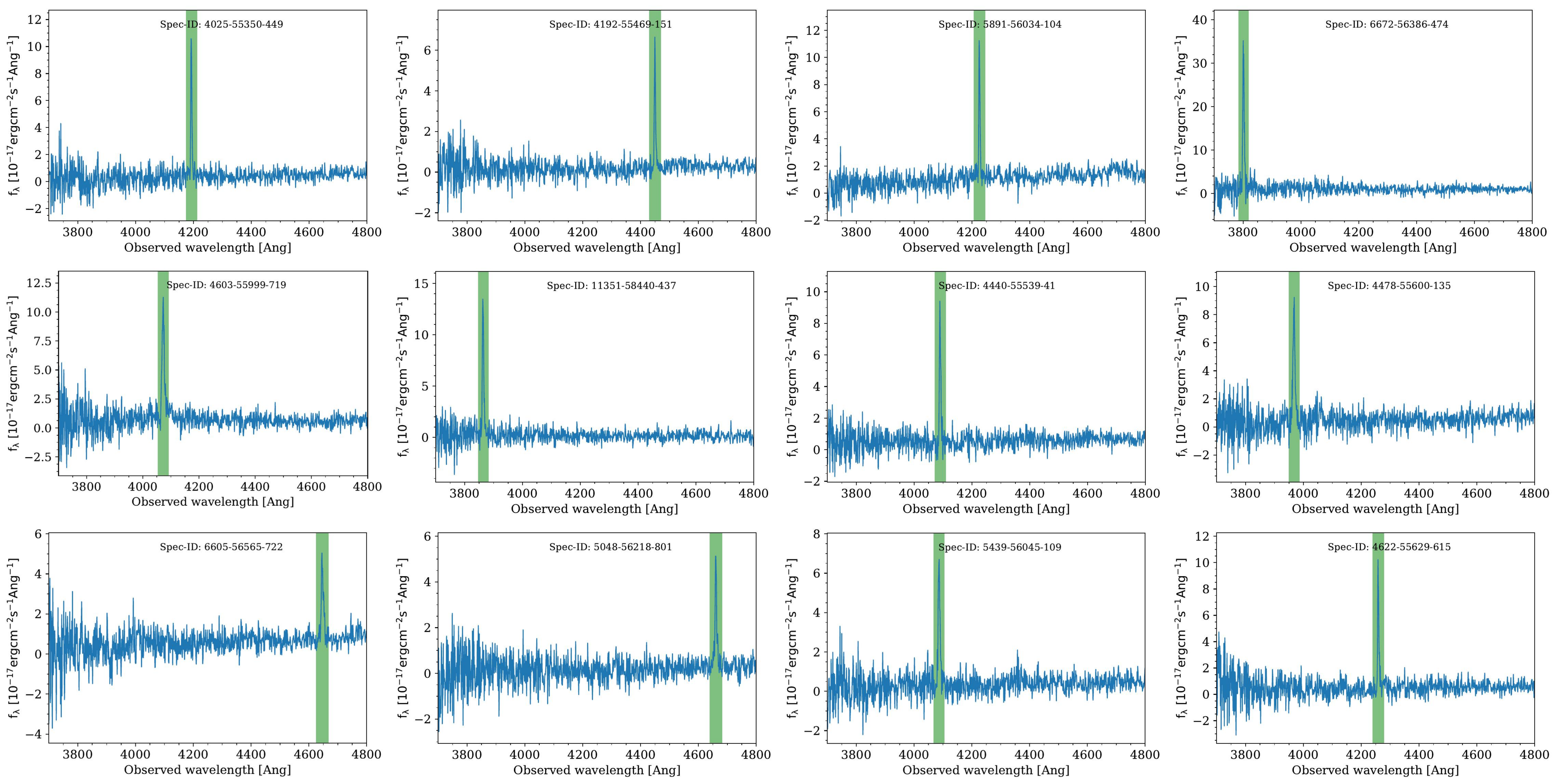}
    \caption{Spectra of 12 single-peaked candidates are shown here for demonstration. The emission lines identified as $\rm Ly\alpha$ are marked by shaded green. The text in each panel shows the SDSS `plate-mjd-fiber.' More spectra figures are available in the online journal.}  
    \label{fig:single_full}
\end{figure*}

\begin{figure*}
	\includegraphics[width=\textwidth]{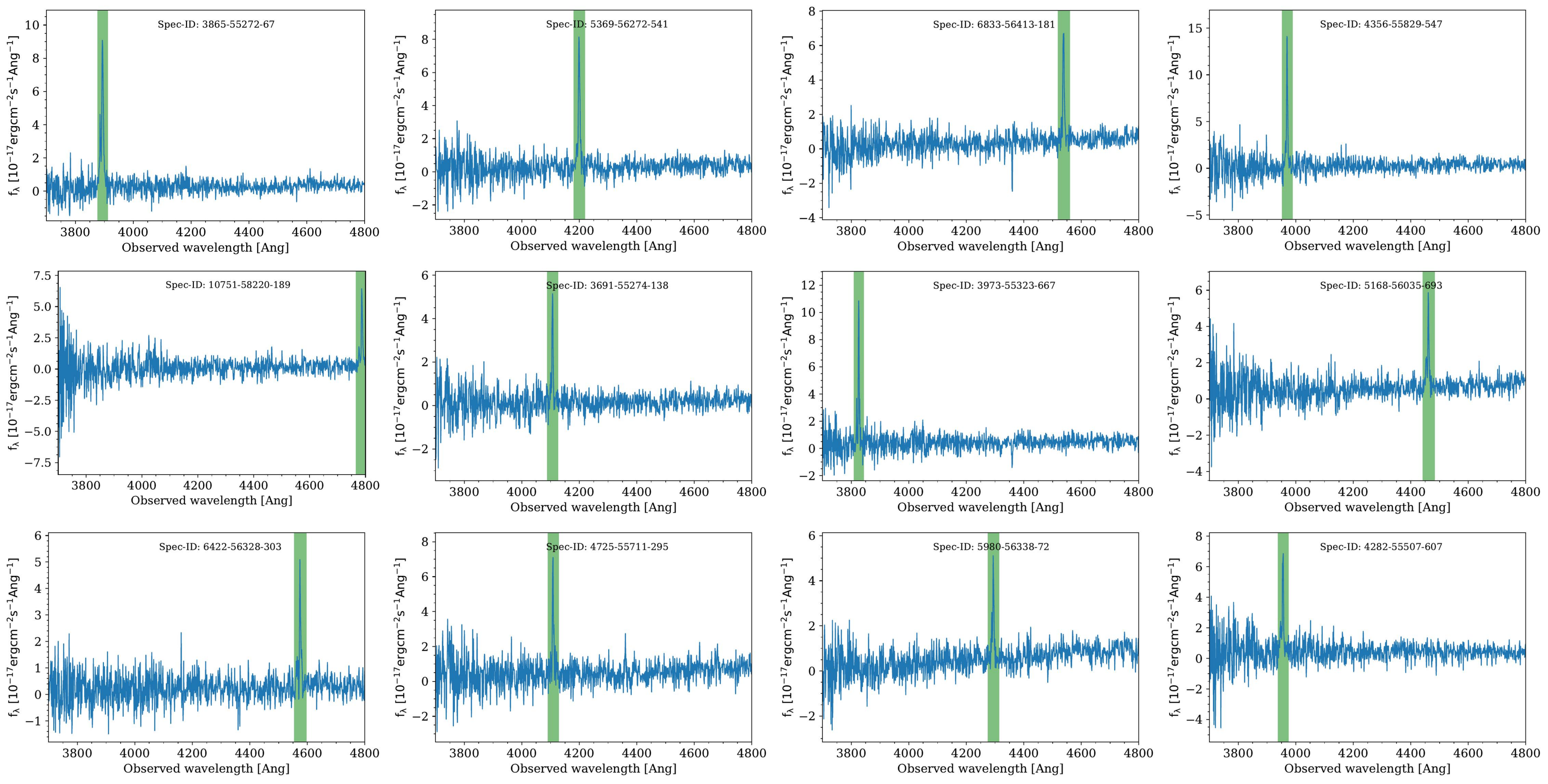}
    \caption{Similar to \ref{fig:single_full}, but for double-peaked candidates.}  
    \label{fig:double_full}
\end{figure*}

\begin{figure*}
	\includegraphics[width=\textwidth]{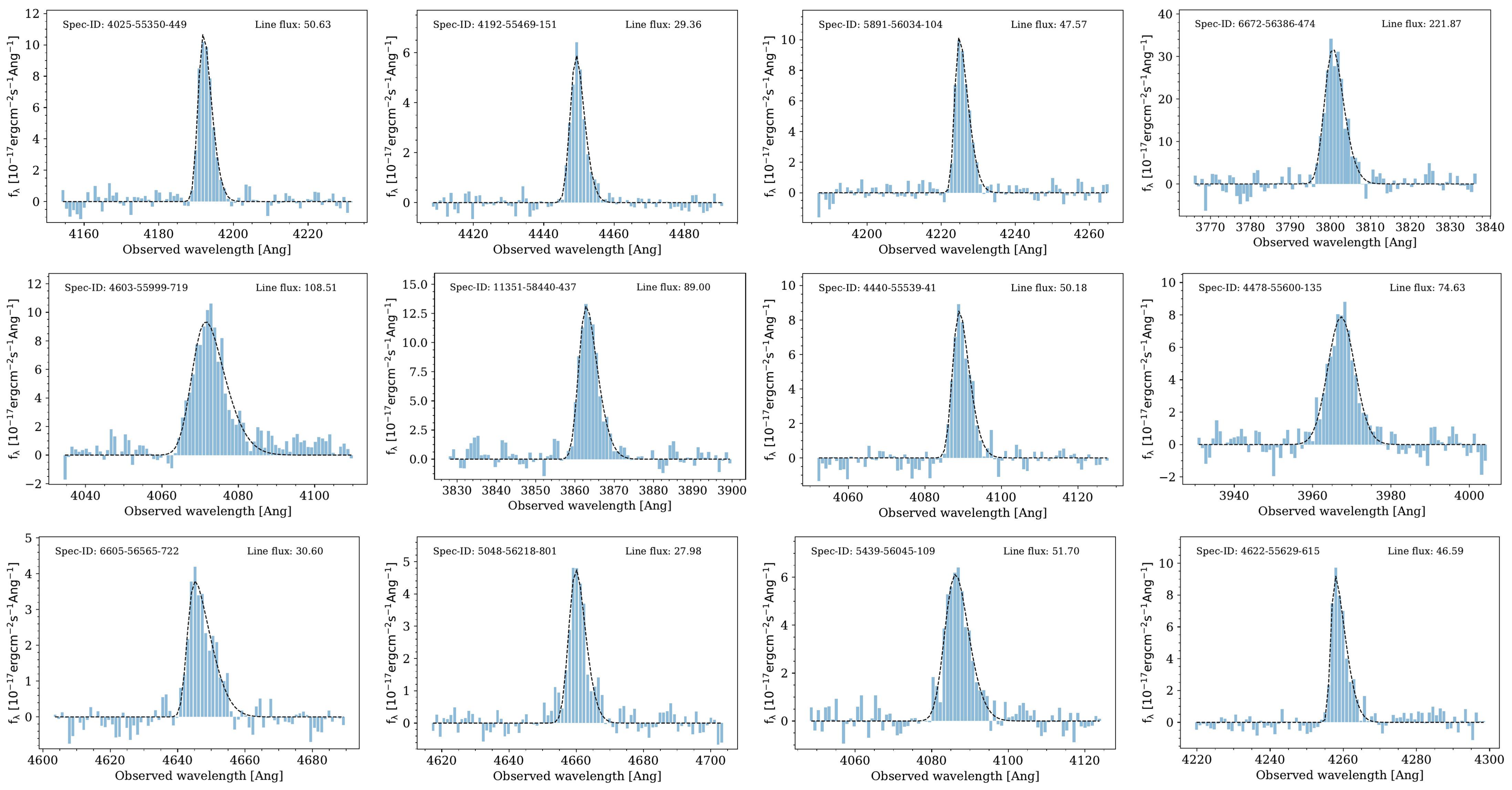}
    \caption{This figure shows zoomed spectra around the $\rm Ly\alpha$ for 12 single-peaked candidates. The blue bars represent residual spectra (observed spectra with the best-fit composite spectra subtracted), and the black dot lines show our best skew-normal fit to the $\rm Ly\alpha$ emission. More zoomed spectra figures are available in the online journal.}
    \label{fig:single_zoom}
\end{figure*}

\begin{figure*}
	\includegraphics[width=\textwidth]{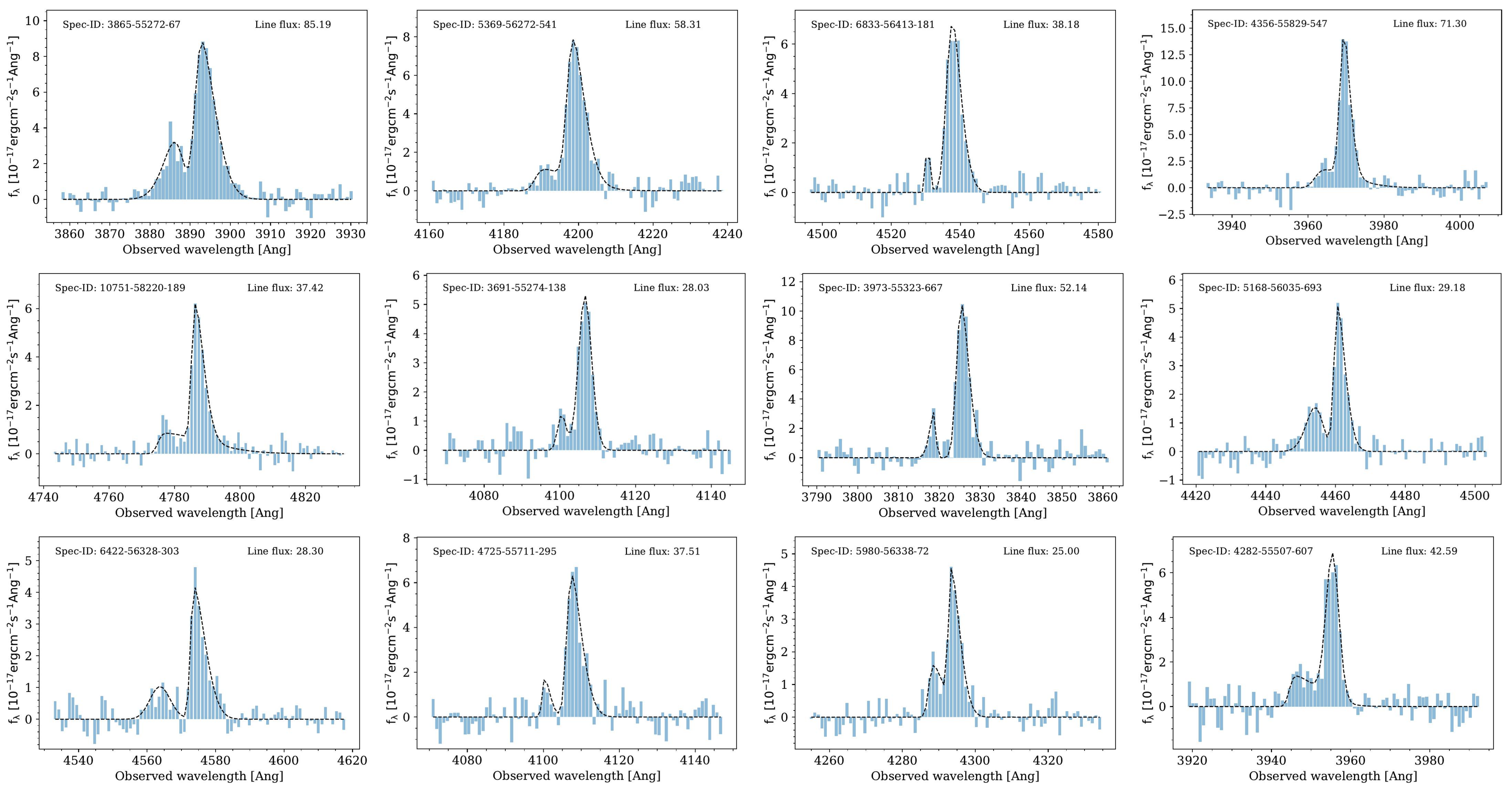}
    \caption{Similar to \ref{fig:single_zoom}, but for double-peaked candidates.}
    \label{fig:double_zoom}
\end{figure*}

In Figures \ref{fig:single_full} and  \ref{fig:double_full}, we show the spectra in wavelength range $3700<\lambda<4800$ \AA\ for our single-peaked and double-peaked candidates, respectively. The shadow regions mark the location of the $\rm Ly\alpha$ emissions. To show the fine structure of $\rm Ly\alpha$ emissions more clearly, we present the zoom in views around the $\rm Ly\alpha$ emission for single-peaked and double-peaked candidates in Figures \ref{fig:single_zoom} and \ref{fig:double_zoom}, respectively. The blue bars in Figures \ref{fig:single_zoom} and \ref{fig:double_zoom} show the raw residual spectra of our lensed LAE candidates, 
%the line peaks in which are modeled by the skew-normal profile with the formula in %\citet{Vivian2015} \citep[see also,][]{Mallery2012},
%%
%\begin{equation}
%\begin{split}
%flux = A e^{-0.5\times((\lambda-x)/\omega)^2}  %[\int_{-\infty}^{s(\lambda-x)/\omega}\exp(-t^2/2)dt] + c \,,
%\end{split}
%\label{eq1}
%\end{equation}
%%
%where $s$ is the skewness. We can calculate the line centre with $\lambda_0 = x %+\omega\delta\sqrt{2/\pi}$ and the line width $\sigma = \omega\sqrt{1 - 2\delta^2/\pi}$, $\delta %= s / \sqrt{1+s^2}$. 
To quantify the $\rm Ly\alpha$ line morphology, we fit the raw-residual spectra with the skew-normal model shown in Equation \eqref{eq0}, using the Nested-Sampling tool -- \textsc{pymultinest} \citep{Buchner2014}, which is a python wrapper of the Fortran code -- \textsc{Multnest} \citep{Feroz2009}. The best-fit values of skewness, width and the total line flux are shown in Table \ref{tab:single_tab} and Table \ref{tab:double_tab}.

\begin{figure*}
	\includegraphics[width=\textwidth]{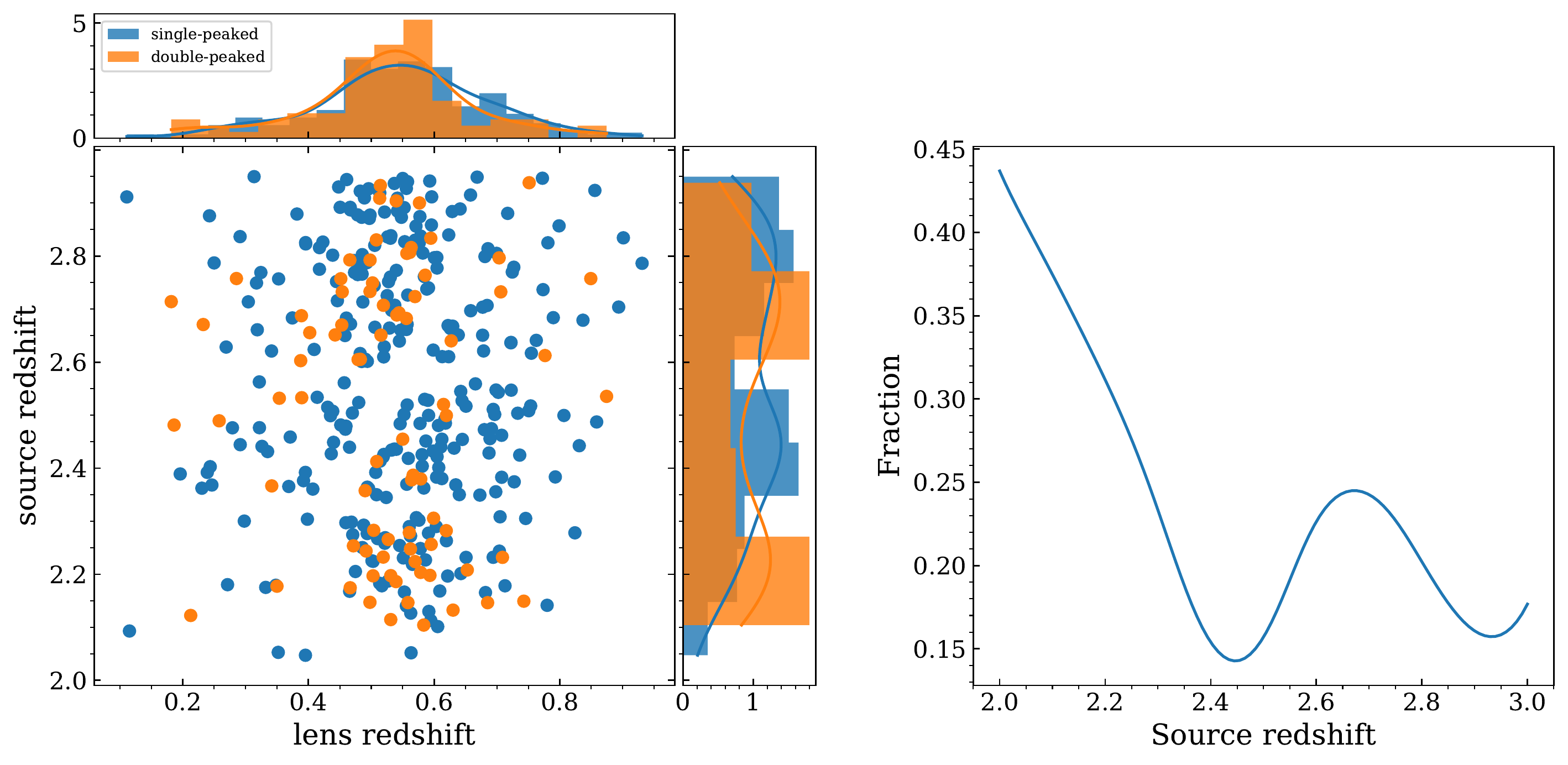}
    \caption{The redshift distribution of lens/source galaxies in our sample. Left: The bottom-left panel shows the redshift distribution of lens/source galaxies in the 2d-plane via scatter-diagram. Double-peaked candidates are marked with orange dots, while blue dots represent single-peaked candidates. The orange and blue histograms in the top panel are the 1d marginalized Probability Density Function (PDF) of lens redshift for double-peaked and single-peaked candidates, respectively. The smooth version of PDF estimated via Gaussian Kernel Density Estimation (KDE) assuming the "Scott" bandwidth scenario is shown by the orange and blue lines, where orange represents single-peaked candidates and blue indicates double-peaked candidates. Similarly, the 1d marginalized PDF of source redshift for double-peaked and single-peaked candidates are shown in the right panel. Right: The fraction of double-peaked candidates as a function of redshift, which is evaluated from the smooth PDF of source redshift using the Gaussian KDE.}
    \label{fig:z_dist}
\end{figure*}

On the left part in Figure \ref{fig:z_dist}, we show the redshift distribution of the lens and the source in our final catalogue. The histogram in the top and right panel show the 1d marginalized probability density function (PDF) for the lens redshift and the source redshift respectively.  We perform the Gaussian kernel density estimation (KDE) for each redshift histogram to derive the smoothed distribution of the redshifts. The lenses in our sample distribute in redshift range 0.2<z<1.0, which peaks at z=0.55, while the sources distribute in redshift range 2<z<3 and seems to have double peaks for both single-peaked and double-peaked samples. On average we find that about $20\%$ of our lensed LAE candidates have the double-peaked line profile; the fraction varies significantly with redshift, from $\sim$45\% at z=2 to $\sim$10\% at redshift z=3, as shown by the right part in Figure \ref{fig:z_dist}. These values agrees broadly with previous studies for field LAEs, which show the fraction of double-peaked LAEs is around 20\%-50\% \citep{Yamada2012,trainor2015, sobral2018, kulas12}.

\begin{figure*}
	\includegraphics[width=\textwidth]{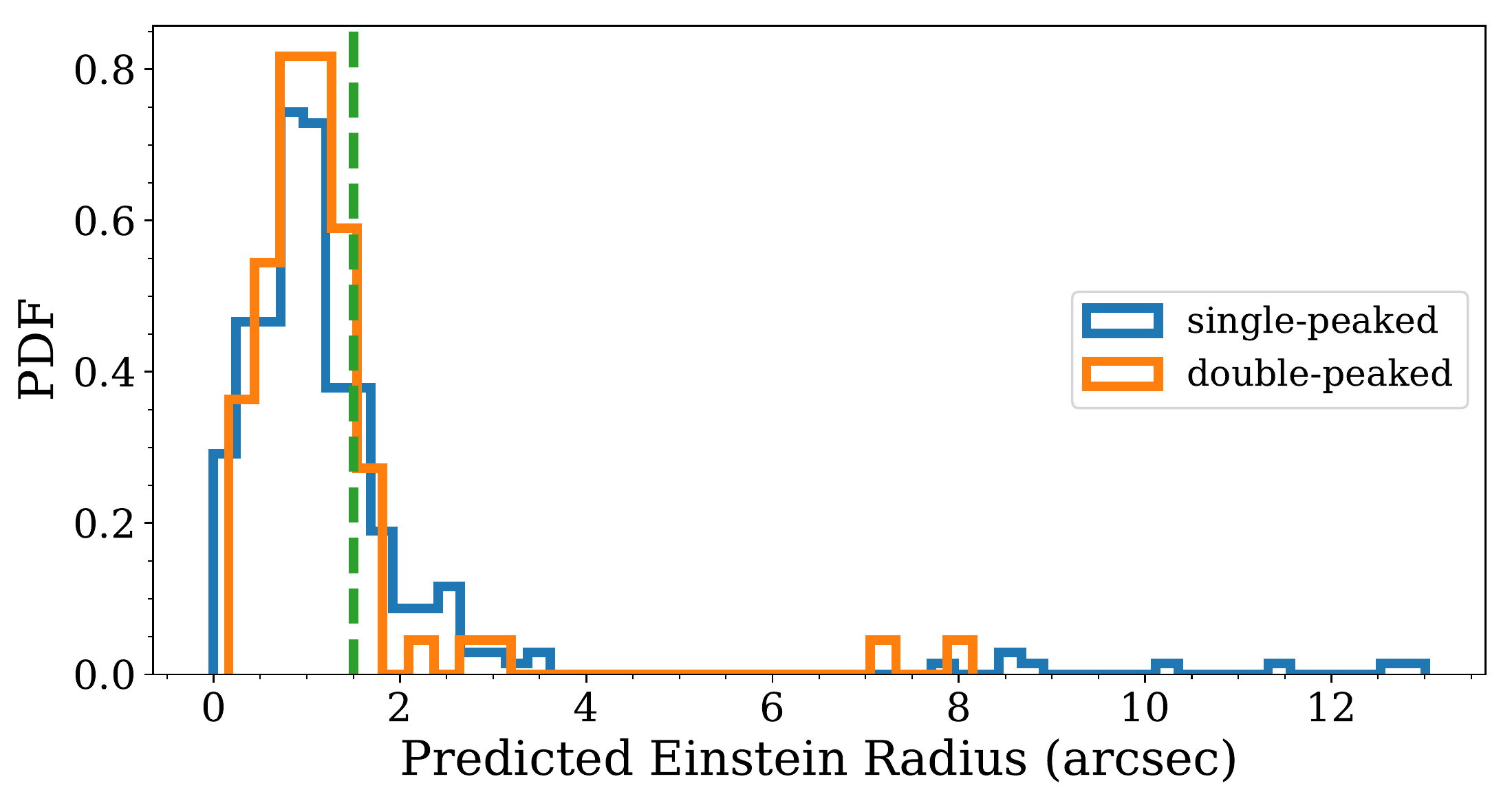}
    \caption{The Probability Density Distribution (PDF) of the predicted Einstein radii assuming the SIE model. Two types of lensed LAEs are shown by different colors. The green dashed line marks the radius of eBOSS fiber (1.5\arcsec). Note that the eBOSS pipeline sometimes generates unreliable estimations for velocity dispersion of target galaxies. Thus some systems in our catalog may have anomalously large velocity dispersions and unreliably large Einstein radii.}
    \label{fig:ein_r}
\end{figure*}

\citet{Bolton2008} empirically show that the probability of a candidate being a real lensing system significantly increases as the predicted Einstein radius becomes larger. To help future follow-up observations,  we calculate the predicted Einstein radius ($\rm \theta_E$) of our candidates, assuming the lens follows a singular isothermal ellipsoid model \citep[SIE,][]{Kormann1994},
\begin{equation}
     b_{sie}=4\pi\sigma_{v}^2/c^2D_{LS}/D_S \,,
\end{equation}
where $\sigma_v$, $D_{LS}$, and $D_{S}$ are the velocity dispersion of the lens galaxy, the angular distance between the lens and source, and the angular distance between the observer and source. The $\rm \theta_E$ for each candidate is also listed in Tables~\ref{tab:single_tab} and \ref{tab:double_tab}. The blue line in Figure \ref{fig:ein_r} shows the probability density distribution of $\rm \theta_E$ for single-peaked candidates, which follows the same distribution as the double-peaked type, which implies single-peaked and double-peaked LAEs are lensed in a similar way. Lensing doesn't introduce additional selection biases to different types of LAE sources, any statistical difference between different types of LAEs inferred from our lensed samples should also exist in the unlensed case.
%{\color{green}which suggests that there is no selection bias for the lenses.}

\begin{figure*}
	\includegraphics[width=\textwidth]{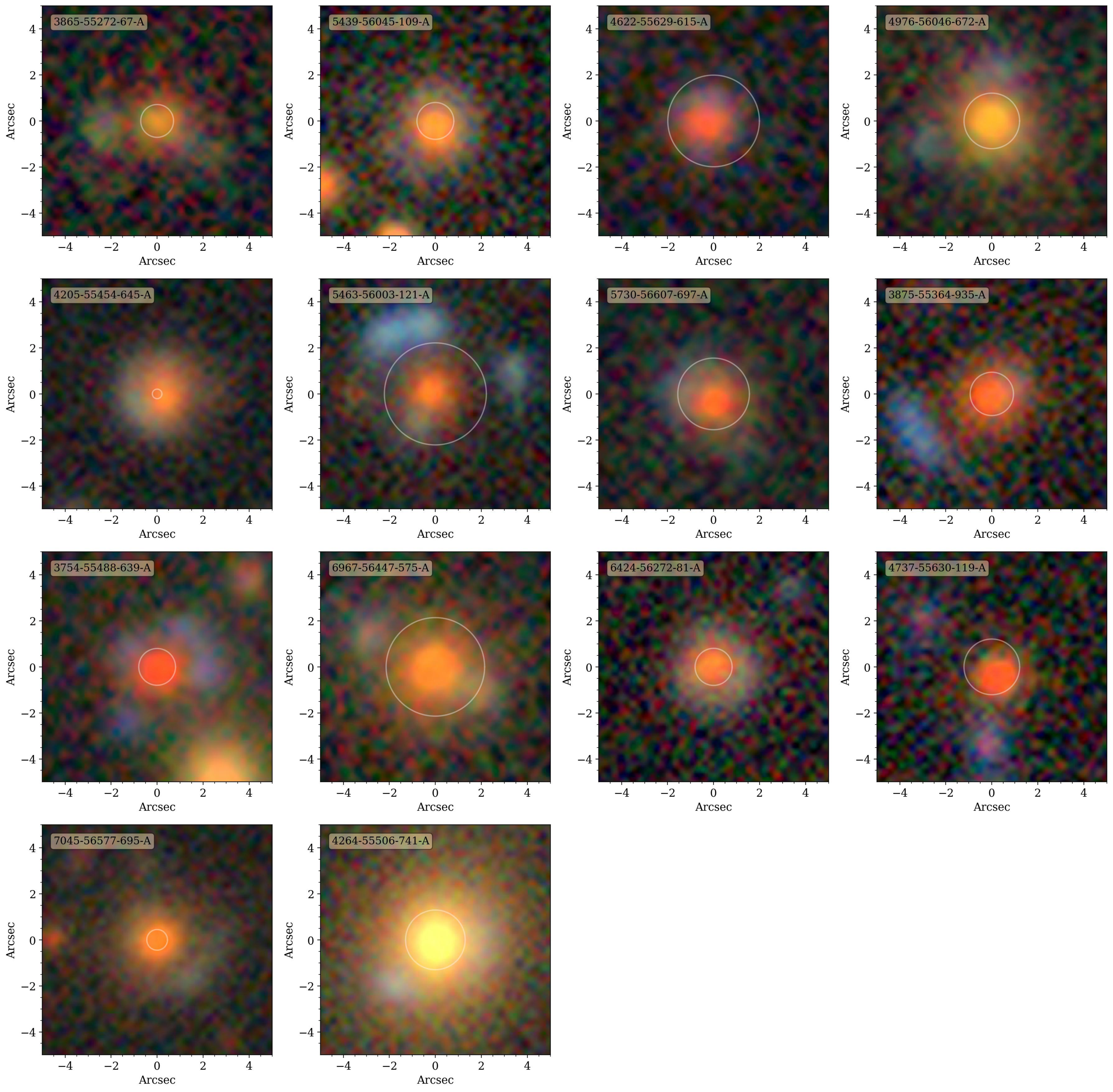}
    \caption{An atlas that includes all the cut-out images of grade-A candidates. Images are taken from the DECaLs survey. The text label in each panel represents the `plate-mjd-fiber-grade.' where the grade is evaluated according to the criterion defined in Section \ref{sec:res}. White circles in each panel show the predicted Einstein radius calculated from the velocity dispersion under the SIE lens assumption.} 
    \label{fig:grade_A}
\end{figure*}

We further check whether our candidates already exhibit some lensed features in the images from 
the DESI Legacy Imaging Surveys \citep{DesiImaging}
%the DECam Legacy Survey (DECaLS) sky-survey 
\footnote{webpage: http://legacysurvey.org/}. We classify our candidates into grade A, B, C by visual inspection. Our classification criteria are as follow:
\begin{enumerate}
\item{Grade A: Multiple images or an arc with a counter-image surrounding the main lens is observed. The lensed feature and main lens have different colors. The Einstein radius inferred from the image is smaller than 5$\arcsec$.}
\item{Grade B: A secondary image or an arc of a different color from the lens object is observed, but no obvious counter-image can be observed.}
\item{Grade C: No discernable lensing feature exists.}
\end{enumerate}
Figure \ref{fig:grade_A} shows all our grade-A candidates. A full atlas that includes all our candidates is available in the online journal. In total, we have 14 grade-A (4 new), 25 grade-B (12 new), and 322 grade-C (213 new) candidates.

\section{Discussion and Summary}

In this work, we present the LESSER strong lensing catalog, which is constructed through a systematic search for galaxy-LAE lensing systems from the eBOSS database. The catalog including 361 candidate lensing systems. The lens galaxies of these systems are LRGs within the redshift range $0.4<z<0.8$, and the source galaxies are LAEs at redshift $2<z<3$. We examine the line profiles of these candidate LAEs and classify them into two subsamples, the ones with single-peaked line profiles (281) and those with double-peaked line profiles (80). 

Our lensed LAE candidates with different types of $\rm Ly\alpha$ emission could be potentially beneficial for the study of both background sources and foreground lens galaxies with future follow-up observations. We describe them in detail as follows.

\subsection{Science of the background source}

A typical lens in our sample can provide a magnification of $\sim10$, boosting both the total flux and the apparent size of the background LAEs. With future high-resolution imaging follow-up, these lenses can offer new possibilities to study the LAEs in detail.  The science potential of the lensed LAEs has been demonstrated in previous work of \citet{Shu2016a}, who obtained the morphology of LAEs down to the scale of $\sim100$ pc region using follow-up Hubble imaging \citep[see also][]{Cornachione2018}, which cannot be achieved outside the strong lensing systems. Base on the same data set, \citet{Ritondale2019a}  built the non-parametric models of the background LAEs, and found these clumps are usually very elliptical, revealing typical configuration of star-formation disks. And this phenomenon may indicate that the escape rate of $\rm Ly\alpha$ photon is irrelevant with the inclination of the star-formation disk. Thus, some anisotropic escape model for LAEs (such as \cite{Verhamme2012}) may not be favored.

One intriguing question of LAEs is why they have different line profiles. Although many theoretical effort has been made to explain the diversity of Ly$\alpha$ emission in LAEs \citep{Hayashino2004,Verhamme2006,Verhamme2012,Dijkstra2012,Orsi2012,Gronke2016}, no consensus has emerged, and more observational evidence is needed to discriminate different theoretical models. If high resolution imaging can be obtained for our sample, they will form a promising lab for this purpose.

\subsection{Application for foreground lens science}

Massive Early-type elliptical Galaxies (ETGs) are the end-product of hierachical galaxy formation. Strong gravitational lensing system can constrain the inner density distributions of the ETGs, which contains crucial information of the galaxy evolution process. 
\citet{Koopmans2006} found the inner mass density slope of lens galaxies do not significantly evolve with redshift by joint gravitational lensing and stellar-dynamical analysis of 15 massive field early-type galaxies selected from the Sloan Lens ACS (SLACS) Survey. They found $d\langle\gamma\rangle/dz=-0.23\pm0.16$ , where the $\langle\gamma\rangle$ represents the inner density slope of total mass. Recently, \citet{chenyun2019} utilized the most comprehensive compiled lens dataset to measure the evolution of the galaxy density profile. By combining the joint gravitational lensing and stellar-dynamical analysis of 161 galaxy-scale strong gravitational lensing systems from the LSD survey \citep{Treu04}, the SL2S survey \citep{Sonnenfeld13}, the SLACS survey, the S4TM survey, the BELLS survey, and the BELLS GALLERY survey, they measured the $d\langle\gamma\rangle/dz=-0.218_{-0.087}^{+0.089}$. This result indicates a mild trend that the galaxies at a lower redshift have steeper slopes. Similar conclusions were also made by several other works \citep{Koopmans2009,Ruff11,Sonnenfeld13,Ruili18}. While the above works have claimed only weak evolution for the total mass density slope, \citet{Bolton12} analyzed the SLACS and BELLS lenses over the redshift interval $z \approx 0.1-0.6$, and they found a stronger evolution trend, with magnitude $d\langle\gamma\rangle/dz=-0.60\pm0.15$. On the other hand, a recent simulation analysis based on the IllustrisTNG project gives a different picture, which claimed that the density slope remains almost constant from $z=1$ to $z=0$, aside from a residual trend of becoming shallower towards $z = 0$ \citep{Yunchong19}. Our lens sample can be complimentary to previous studies in the sense that a significant fraction ($\sim$10\%) of our lenses locate at $z>0.7$, which rarely appears in previous BELLS and BELLS GALLERY catalog. These new $z>0.7$ lenses will allow us to extend the strong-lensing measurement of the galaxy inner density slope to higher redshift and obtain a more comprehensive understanding of the evolution of the density slope of ETGs.

Another important application of the lensed LAEs is the searching of low mass dark matter haloes and subhaloes through the flux distortions that they induce on the lensed images. For galaxy lenses, with high-resolution imaging observations from a ground-based telescope with Laser-AO, the mass limit of detection for haloes (subhaloes) can be as low as $\sim 10^8 M_{\odot}$
%$\sim 10^7 M_{\odot}$ 
\citep{Vegetti2012}. Recent theoretical investigations have shown that a sample of $\sim 20$ with detection limit of $\sim 10^8$ $M_{\odot}$ can distinguish cold dark matter model from the warm dark matter model which are made of 7 keV sterile neutrinos whose decay can explain the observed 3.5 kev X-ray emission lines in clusters and galaxies \citep{ranli2016, ranli2017, Boyarsky2014, Bulbul2014}.

The flux perturbation induced by the dark matter clumps in lens galaxy is proportional to the product of brightness gradient of the source galaxy and the gradient of the subhalo potential perturbation. For a given detection precision of flux distortion, the higher the source gradient is, the lower the lens potential perturbation can be detected \citep{Koopmans2005}. The LAEs are usually made of multiple compact clumps with sharp brightness profile. Therefore, our candidates are promising places for the search of low mass dark matter subhaloes \citep{Shu2016a, Ritondale2019b}, if high-resolution imaging follow-ups are obtained.

%------------------------------------figure

%----------------------------------------------full--spec

%----------------------------------------------full--spec

%----------------------------------------------zoomed--spec

%----------------------------------------------zoomed--spec

%----------------------------------grade image

\begin{table*}
	\centering
	\caption{Catalog of single-peaked candidates . From left to right: the SDSS plate-fiber-mjd, right ascension and declination of lens galaxies, redshifts of lens galaxies and background LAEs, i-band SDSS cmodel magnitude of lens galaxies, the line flux, width and skewness of $\rm Ly\alpha$ emission quantified by the skew-normal fit, the flux and width are in units of $\rm 10^{-17}\,erg\,cm^{-2}\,s^{-1}$ and \AA, respectively. The velocity dispersions of lens galaxies in units of $\rm km/s$ and associated predicted Einstein radii in units of arcsecond. The last column gives the grade of our candidates based on color images from the DECaLs survey. Candidates already found by the BELLS GALLERY project are marked with star symbols in the last column. We only list part of the catalog here for conciseness, a full catalog is available in the online journal.}
	\label{tab:single_tab}

%Note: Candidate 9354-57806-49 is repetitive to 3754-55488-639 since this object is observed twice in the eBOSS survey.
%--------------------------------------------------------------------------
\begin{tabular}{cccccccccccc}
\hline
name & ra & dec & $z_l$ & $z_s$ & $mag_i$ & $F_{lya}$ & width & skewness & $\rm \sigma_{eBOSS}$ & $\theta_E$ & grade \\
\hline
4025-55350-449 & 14:31:33.30 & -00:44:54.17 & 0.44 & 2.45 & 19.87 & 50.63 & 2.10 & 1.97 & 136.10 & 0.38 & $\rm C^{\star}$ \\
4192-55469-151 & 21:20:47.64 & -00:09:11.33 & 0.55 & 2.66 & 19.96 & 29.36 & 2.21 & 1.60 & 232.59 & 1.03 & $\rm C^{\star}$ \\
5891-56034-104 & 11:39:39.00 & +16:51:58.52 & 0.28 & 2.48 & 17.89 & 47.57 & 2.09 & 2.82 & 206.90 & 1.00 & $\rm C^{\star}$ \\
6672-56386-474 & 12:01:59.03 & +47:43:23.19 & 0.56 & 2.13 & 19.30 & 221.87 & 3.10 & 1.45 & 226.79 & 0.90 & $\rm C^{\star}$ \\
4603-55999-719 & 09:07:34.40 & +42:55:28.56 & 0.64 & 2.35 & 19.93 & 108.51 & 5.24 & 1.69 & 139.33 & 0.33 & B \\
11351-58440-437 & 10:13:23.75 & +28:12:00.10 & 0.71 & 2.18 & 20.87 & 89.00 & 3.05 & 1.93 & 167.30 & 0.43 & C \\
4440-55539-41 & 07:42:49.68 & +33:41:48.99 & 0.49 & 2.36 & 19.47 & 50.18 & 2.63 & 1.94 & 249.24 & 1.20 & $\rm C^{\star}$ \\
4478-55600-135 & 08:30:52.17 & +21:28:58.13 & 0.62 & 2.26 & 19.91 & 74.63 & 4.00 & 0.56 & 229.63 & 0.90 & $\rm C^{\star}$ \\
6605-56565-722 & 02:01:21.40 & +32:28:29.67 & 0.40 & 2.82 & 18.33 & 30.60 & 3.68 & 3.15 & 258.23 & 1.44 & $\rm B^{\star}$ \\
5048-56218-801 & 22:28:25.77 & +12:05:03.95 & 0.53 & 2.83 & 19.90 & 27.98 & 2.62 & 1.54 & 241.72 & 1.14 & $\rm C^{\star}$ \\
5439-56045-109 & 13:44:08.02 & +16:21:27.99 & 0.50 & 2.36 & 19.29 & 51.70 & 3.77 & 1.38 & 204.02 & 0.80 & $\rm A^{\star}$ \\
4622-55629-615 & 11:10:40.43 & +36:49:24.39 & 0.73 & 2.50 & 19.87 & 46.59 & 2.30 & 4.00 & 353.54 & 1.99 & $\rm A^{\star}$ \\
6281-56295-811 & 00:29:27.39 & +25:44:01.78 & 0.59 & 2.45 & 19.22 & 37.75 & 1.88 & 1.85 & 212.65 & 0.81 & $\rm C^{\star}$ \\
6727-56369-967 & 15:16:41.22 & +49:54:40.78 & 0.55 & 2.87 & 19.02 & 29.64 & 2.08 & 1.81 & 270.91 & 1.42 & $\rm C^{\star}$ \\
4976-56046-672 & 15:42:04.99 & +40:19:54.69 & 0.44 & 2.50 & 18.78 & 42.53 & 3.68 & 3.46 & 242.52 & 1.21 & $\rm A^{\star}$ \\
6147-56239-163 & 23:17:31.62 & +11:54:50.92 & 0.54 & 2.64 & 19.63 & 39.67 & 6.15 & 1.53 & 272.55 & 1.41 & $\rm C^{\star}$ \\
4399-55811-149 & 02:37:40.64 & -06:41:12.99 & 0.49 & 2.25 & 19.23 & 38.16 & 2.24 & 1.90 & 0.00 & 0.00 & $\rm C^{\star}$ \\
5813-56363-341 & 09:18:07.86 & +45:18:56.78 & 0.52 & 2.34 & 19.35 & 44.35 & 2.35 & 1.89 & 80.08 & 0.12 & $\rm B^{\star}$ \\
4615-55618-399 & 11:31:10.98 & +35:50:20.72 & 0.47 & 2.30 & 19.76 & 37.94 & 2.87 & 1.04 & 102.04 & 0.20 & $\rm C^{\star}$ \\
6497-56329-271 & 14:13:58.21 & +29:32:40.41 & 0.45 & 2.72 & 19.41 & 17.28 & 1.76 & 1.34 & 158.87 & 0.52 & $\rm C^{\star}$ \\
4315-55503-703 & 01:13:00.58 & +02:50:46.21 & 0.62 & 2.61 & 19.46 & 29.16 & 2.64 & 2.99 & 697.90 & 8.65 & $\rm B^{\star}$ \\
4205-55454-645 & 22:45:05.93 & +00:40:18.35 & 0.70 & 2.54 & 19.57 & 27.96 & 4.69 & 1.98 & 112.39 & 0.21 & $\rm A^{\star}$ \\
6435-56341-855 & 11:10:27.11 & +28:08:38.46 & 0.61 & 2.40 & 19.91 & 33.31 & 2.74 & 2.89 & 193.52 & 0.66 & $\rm C^{\star}$ \\
6659-56607-751 & 10:29:01.23 & +47:28:22.64 & 0.57 & 2.83 & 19.11 & 18.76 & 2.22 & 2.34 & 213.10 & 0.86 & $\rm C^{\star}$ \\
5463-56003-121 & 14:29:54.80 & +12:02:35.58 & 0.55 & 2.83 & 19.60 & 24.19 & 3.59 & 2.98 & 339.91 & 2.22 & $\rm A^{\star}$ \\
6840-56685-377 & 12:41:44.24 & +60:24:41.06 & 0.53 & 2.84 & 19.19 & 20.18 & 2.30 & 2.04 & 270.38 & 1.43 & $\rm C^{\star}$ \\
4209-55478-171 & 23:14:18.03 & -01:10:27.03 & 0.52 & 2.27 & 19.85 & 25.44 & 2.01 & 1.42 & 279.95 & 1.46 & $\rm C^{\star}$ \\
5730-56607-697 & 09:18:59.21 & +51:04:52.60 & 0.58 & 2.40 & 19.71 & 26.27 & 2.30 & 8.69 & 294.41 & 1.56 & $\rm A^{\star}$ \\
3802-55514-227 & 08:05:30.94 & +38:09:23.36 & 0.56 & 2.27 & 19.00 & 33.51 & 2.29 & 1.92 & 304.99 & 1.67 & C \\
3786-55563-480 & 10:43:48.35 & -02:12:36.06 & 0.77 & 2.74 & 19.92 & 17.22 & 2.11 & 3.16 & 850.00 & 11.51 & C \\
6521-56537-324 & 23:52:12.72 & +21:45:26.18 & 0.51 & 2.92 & 19.74 & 15.63 & 2.10 & 1.82 & 196.01 & 0.76 & $\rm C^{\star}$ \\
5176-56221-583 & 08:43:00.14 & +19:47:42.63 & 0.70 & 2.55 & 19.81 & 17.47 & 1.79 & 2.68 & 384.25 & 2.44 & $\rm C^{\star}$ \\
3952-55330-944 & 15:46:11.14 & +28:05:42.59 & 0.30 & 2.71 & 17.95 & 21.08 & 2.11 & 3.95 & 215.58 & 1.07 & $\rm C^{\star}$ \\
5984-56337-186 & 12:41:32.92 & +25:32:58.91 & 0.54 & 2.71 & 19.39 & 26.98 & 3.20 & 1.38 & 193.60 & 0.72 & $\rm C^{\star}$ \\
3683-55178-662 & 07:59:30.25 & +44:12:45.56 & 0.47 & 2.89 & 19.76 & 11.64 & 1.84 & 1.27 & 192.75 & 0.76 & $\rm C^{\star}$ \\
11316-58396-95 & 21:15:36.90 & +04:52:46.79 & 0.89 & 2.70 & 21.30 & 16.38 & 1.97 & 1.39 & 89.45 & 0.11 & C \\
5068-55749-829 & 22:13:21.15 & +09:02:41.06 & 0.63 & 2.67 & 19.80 & 17.82 & 1.99 & 1.36 & 224.06 & 0.89 & $\rm C^{\star}$ \\
3673-55178-795 & 07:48:31.92 & +47:10:14.40 & 0.44 & 2.51 & 19.17 & 17.86 & 2.91 & 2.40 & 139.77 & 0.40 & $\rm C^{\star}$ \\
5366-55958-259 & 11:15:44.57 & +08:47:17.16 & 0.55 & 2.89 & 19.55 & 16.40 & 2.19 & 2.07 & 235.26 & 1.07 & $\rm C^{\star}$ \\
6725-56390-772 & 14:43:17.84 & +51:07:21.03 & 0.55 & 2.95 & 18.73 & 16.33 & 2.82 & 9.40 & 169.93 & 0.56 & $\rm B^{\star}$ \\
4619-55599-680 & 11:25:58.59 & +33:56:47.11 & 0.48 & 2.92 & 19.46 & 15.45 & 1.98 & 0.15 & 196.24 & 0.78 & $\rm C^{\star}$ \\
3929-55335-927 & 16:11:57.06 & +22:14:40.00 & 0.47 & 2.27 & 19.85 & 24.12 & 2.54 & 0.80 & 336.16 & 2.20 & C \\
4091-55498-227 & 21:48:43.04 & +04:44:36.11 & 0.66 & 2.70 & 19.85 & 10.73 & 1.63 & 1.00 & 375.63 & 2.46 & $\rm C^{\star}$ \\
4236-55479-117 & 02:17:42.02 & -00:22:06.47 & 0.61 & 2.48 & 19.47 & 14.71 & 2.80 & 1.85 & 218.18 & 0.84 & $\rm C^{\star}$ \\
7162-56605-731 & 01:36:44.88 & -06:46:56.69 & 0.67 & 2.95 & 19.90 & 12.04 & 2.12 & 3.08 & 224.43 & 0.89 & $\rm C^{\star}$ \\
8308-57417-911 & 07:56:50.15 & +51:06:48.34 & 0.51 & 2.27 & 20.41 & 24.16 & 4.32 & 0.13 & 308.25 & 1.79 & C \\
5069-56211-141 & 22:21:51.54 & +07:24:00.24 & 0.75 & 2.51 & 19.65 & 14.69 & 1.79 & 1.41 & 308.02 & 1.49 & $\rm C^{\star}$ \\
4030-55634-117 & 14:18:15.73 & +01:58:32.23 & 0.56 & 2.14 & 19.35 & 33.88 & 2.96 & 2.61 & 219.28 & 0.85 & $\rm C^{\star}$ \\
8735-58133-17 & 02:04:57.62 & +06:03:59.14 & 0.72 & 2.77 & 19.99 & 33.16 & 2.42 & 2.09 & 270.16 & 1.22 & C \\
6297-56218-817 & 22:33:33.17 & +27:20:37.54 & 0.53 & 2.70 & 19.86 & 14.67 & 2.19 & 1.94 & 195.12 & 0.73 & $\rm C^{\star}$ \\
5314-55952-733 & 09:35:51.05 & +08:51:30.31 & 0.46 & 2.56 & 19.55 & 18.56 & 1.74 & 1.32 & 293.60 & 1.75 & $\rm C^{\star}$ \\
5712-56602-365 & 09:15:26.16 & +58:51:55.32 & 0.78 & 2.83 & 19.82 & 13.19 & 2.01 & 2.08 & 318.15 & 1.62 & $\rm C^{\star}$ \\
3960-55663-695 & 15:21:05.07 & +25:41:10.01 & 0.60 & 2.43 & 19.65 & 21.22 & 2.59 & 5.56 & 180.85 & 0.58 & C \\
10740-58223-219 & 15:46:12.04 & +34:52:54.49 & 0.93 & 2.79 & 20.83 & 17.64 & 2.74 & 6.79 & 850.00 & 10.17 & C \\
8849-57457-315 & 11:17:51.25 & +36:23:43.68 & 0.68 & 2.47 & 20.47 & 27.77 & 5.27 & 0.68 & 245.18 & 1.00 & B \\
3946-55659-954 & 15:54:58.12 & +26:46:30.73 & 0.29 & 2.44 & 17.79 & 15.78 & 2.24 & 0.56 & 208.70 & 1.00 & $\rm C^{\star}$ \\
5896-56047-358 & 14:10:26.40 & +21:28:52.30 & 0.31 & 2.95 & 18.18 & 15.15 & 2.87 & 3.81 & 137.85 & 0.44 & $\rm C^{\star}$ \\
%3924-55332-559 & 16:03:24.84 & +20:31:08.47 & 0.55 & 2.17 & 19.76 & 49.38 & 3.23 & 1.81 & 89.48 & 0.14 & $\rm B^{\star}$ \\
%4418-55862-165 & 00:33:38.97 & +04:27:34.94 & 0.53 & 2.19 & 19.66 & 30.06 & 3.29 & 3.43 & 201.63 & 0.74 & $\rm C^{\star}$ \\
%5984-56337-923 & 12:43:00.17 & +26:23:12.57 & 0.45 & 2.89 & 18.79 & 17.78 & 2.37 & 2.54 & 179.12 & 0.67 & $\rm B^{\star}$ \\
\hline
\end{tabular}
\end{table*}
%------------------------------------------single tab finished

%--------------------------------double tab begin
\begin{table*}
	\centering
	\caption{Similar to Table \ref{tab:single_tab}, but for double-peaked candidates. `width' and `skewness' are the characteristics of the red peak.}
	\label{tab:double_tab}
%---------------------------------------------------------------------------------
\begin{tabular}{cccccccccccc}
\hline
name & ra & dec & $z_l$ & $z_s$ & $mag_i$ & $F_{lya}$ & width & skewness & $\rm \sigma_{eBOSS}$ & $\theta_E$ & grade \\
\hline
3865-55272-67 & 14:43:20.73 & +33:42:12.10 & 0.58 & 2.20 & 19.56 & 85.19 & 3.18 & 2.75 & 201.93 & 0.71 & $\rm A^{\star}$ \\
5369-56272-541 & 11:16:34.56 & +09:15:03.05 & 0.55 & 2.45 & 19.41 & 58.31 & 2.75 & 2.56 & 256.14 & 1.22 & $\rm B^{\star}$ \\
6833-56413-181 & 12:26:56.46 & +54:57:39.05 & 0.50 & 2.73 & 18.88 & 38.18 & 2.29 & 1.72 & 237.51 & 1.12 & $\rm C^{\star}$ \\
4356-55829-547 & 23:42:48.68 & -01:20:32.54 & 0.53 & 2.27 & 19.58 & 71.30 & 1.65 & 1.83 & 286.28 & 1.52 & $\rm C^{\star}$ \\
10751-58220-189 & 14:51:58.54 & +35:05:44.62 & 0.75 & 2.94 & 20.55 & 37.42 & 1.88 & 3.79 & 258.86 & 1.11 & C \\
3691-55274-138 & 08:15:35.86 & +45:21:33.57 & 0.56 & 2.38 & 20.12 & 28.03 & 1.93 & 0.69 & 309.53 & 1.74 & $\rm C^{\star}$ \\
3973-55323-667 & 12:46:58.13 & +37:27:00.16 & 0.50 & 2.15 & 19.35 & 52.14 & 1.92 & 1.84 & 93.35 & 0.16 & C \\
5168-56035-693 & 15:02:56.13 & +38:20:17.89 & 0.45 & 2.67 & 18.74 & 29.18 & 1.74 & 3.00 & 230.71 & 1.09 & $\rm C^{\star}$ \\
6422-56328-303 & 11:41:54.71 & +22:16:28.89 & 0.59 & 2.76 & 19.73 & 28.30 & 2.32 & 5.22 & 267.86 & 1.33 & $\rm C^{\star}$ \\
4725-55711-295 & 15:58:21.52 & +26:58:29.77 & 0.58 & 2.38 & 19.66 & 37.51 & 2.34 & 2.88 & 285.37 & 1.46 & $\rm C^{\star}$ \\
5980-56338-72 & 12:33:18.42 & +25:23:12.13 & 0.39 & 2.53 & 18.68 & 25.00 & 1.79 & 5.57 & 220.82 & 1.04 & C \\
4282-55507-607 & 23:33:11.12 & +02:23:10.93 & 0.47 & 2.25 & 19.93 & 42.59 & 1.76 & 0.22 & 125.11 & 0.30 & $\rm B^{\star}$ \\
6160-56190-854 & 23:46:17.29 & +09:09:31.44 & 0.18 & 2.71 & 17.36 & 27.63 & 1.61 & 7.59 & 211.96 & 1.14 & $\rm C^{\star}$ \\
3676-55186-623 & 07:48:16.74 & +44:56:02.84 & 0.54 & 2.19 & 19.05 & 42.73 & 2.93 & 1.81 & 206.27 & 0.77 & $\rm C^{\star}$ \\
4473-55589-231 & 07:50:57.45 & +22:10:35.84 & 0.45 & 2.76 & 19.33 & 21.50 & 2.03 & 1.84 & 207.03 & 0.88 & $\rm C^{\star}$ \\
7050-56573-273 & 01:45:26.41 & -05:00:35.59 & 0.57 & 2.72 & 19.95 & 17.73 & 2.31 & 2.25 & 285.54 & 1.53 & $\rm B^{\star}$ \\
7166-56602-783 & 23:53:56.11 & -08:32:38.86 & 0.48 & 2.60 & 19.28 & 16.98 & 2.27 & 2.23 & 236.98 & 1.12 & $\rm C^{\star}$ \\
6817-56455-897 & 13:28:08.68 & +61:57:17.63 & 0.53 & 2.20 & 19.63 & 35.31 & 3.33 & 9.74 & 283.12 & 1.46 & $\rm C^{\star}$ \\
6014-56072-764 & 14:28:23.29 & +23:58:39.22 & 0.47 & 2.79 & 19.37 & 17.94 & 1.75 & 1.74 & 227.36 & 1.06 & $\rm C^{\star}$ \\
5705-56194-259 & 00:59:48.32 & +16:06:29.49 & 0.59 & 2.83 & 19.98 & 13.48 & 1.82 & 2.47 & 203.82 & 0.77 & $\rm C^{\star}$ \\
4657-55591-602 & 01:51:50.89 & +14:30:37.36 & 0.56 & 2.25 & 19.19 & 46.44 & 2.08 & 4.76 & 255.66 & 1.17 & $\rm C^{\star}$ \\
6506-56564-729 & 23:02:17.65 & +30:27:38.29 & 0.56 & 2.81 & 19.79 & 14.94 & 2.00 & 2.19 & 92.72 & 0.16 & $\rm C^{\star}$ \\
4723-56033-445 & 15:28:51.87 & +31:02:33.89 & 0.74 & 2.15 & 19.81 & 64.99 & 3.05 & 2.05 & 428.00 & 2.70 & $\rm B^{\star}$ \\
3606-55182-695 & 01:51:45.36 & -00:00:01.31 & 0.49 & 2.36 & 19.72 & 22.52 & 1.88 & 3.92 & 241.93 & 1.13 & $\rm C^{\star}$ \\
6732-56370-859 & 14:50:01.80 & +48:18:32.21 & 0.52 & 2.23 & 19.47 & 18.10 & 2.13 & 5.82 & 226.33 & 0.95 & C \\
4644-55922-885 & 09:21:33.07 & +36:37:17.40 & 0.56 & 2.68 & 19.91 & 17.10 & 1.84 & 1.59 & 219.14 & 0.91 & $\rm C^{\star}$ \\
5960-56097-209 & 21:50:34.64 & +24:50:16.53 & 0.50 & 2.28 & 19.10 & 23.86 & 2.44 & 1.59 & 201.49 & 0.77 & $\rm C^{\star}$ \\
6459-56273-547 & 09:57:55.84 & +24:09:44.70 & 0.58 & 2.10 & 19.66 & 56.58 & 4.14 & 3.40 & 110.56 & 0.21 & C \\
4707-55653-211 & 13:25:25.47 & +39:39:14.54 & 0.51 & 2.83 & 19.82 & 13.96 & 3.59 & 3.79 & 199.61 & 0.79 & $\rm C^{\star}$ \\
7585-57190-637 & 22:34:03.96 & +22:10:47.60 & 0.85 & 2.76 & 21.07 & 8.11 & 1.36 & 2.34 & 259.16 & 1.01 & C \\
4398-55946-379 & 02:02:41.41 & -06:46:11.33 & 0.50 & 2.75 & 19.40 & 17.74 & 2.35 & 3.96 & 159.53 & 0.50 & $\rm C^{\star}$ \\
4235-55451-195 & 02:09:58.61 & -00:12:32.73 & 0.47 & 2.17 & 19.68 & 19.97 & 1.81 & 4.19 & 144.71 & 0.40 & C \\
7120-56720-417 & 12:29:10.35 & +66:09:24.48 & 0.51 & 2.41 & 19.43 & 41.45 & 2.61 & 8.92 & 196.17 & 0.74 & $\rm C^{\star}$ \\
6205-56187-591 & 00:42:32.96 & +11:31:13.74 & 0.48 & 2.61 & 19.83 & 13.71 & 2.60 & 4.55 & 180.82 & 0.65 & C \\
5004-55711-151 & 17:32:25.89 & +26:42:40.33 & 0.50 & 2.20 & 19.95 & 15.67 & 2.75 & 5.18 & 176.88 & 0.59 & $\rm C^{\star}$ \\
7033-56565-181 & 23:55:25.25 & -05:21:38.63 & 0.49 & 2.24 & 19.29 & 27.11 & 2.37 & 0.95 & 294.36 & 1.65 & $\rm C^{\star}$ \\
3923-55325-932 & 16:00:07.30 & +14:51:12.23 & 0.34 & 2.37 & 18.13 & 18.23 & 3.14 & 4.84 & 193.75 & 0.82 & $\rm C^{\star}$ \\
5359-55953-51 & 11:05:00.96 & +07:37:30.47 & 0.56 & 2.81 & 19.01 & 12.96 & 1.64 & 5.96 & 615.40 & 7.23 & B \\
6497-56329-843 & 14:16:13.97 & +30:14:57.05 & 0.57 & 2.39 & 19.49 & 8.86 & 2.11 & 0.39 & 303.54 & 1.67 & $\rm C^{\star}$ \\
5172-56071-193 & 14:48:17.66 & +39:49:25.15 & 0.40 & 2.66 & 18.30 & 18.89 & 2.57 & 9.18 & 227.95 & 1.11 & C \\
6315-56181-654 & 16:50:03.85 & +49:42:49.82 & 0.39 & 2.69 & 18.14 & 9.42 & 2.32 & 9.95 & 279.80 & 1.69 & $\rm C^{\star}$ \\
6785-56487-647 & 15:45:25.18 & +59:36:25.99 & 0.71 & 2.73 & 19.86 & 12.54 & 1.90 & 7.77 & 154.63 & 0.40 & C \\
3876-55245-223 & 15:07:07.74 & +30:47:56.83 & 0.62 & 2.50 & 19.87 & 11.60 & 2.62 & 5.22 & 155.21 & 0.42 & B \\
4005-55325-923 & 13:05:21.79 & +02:27:24.53 & 0.63 & 2.13 & 19.43 & 30.27 & 3.58 & 4.53 & 206.58 & 0.70 & C \\
6177-56268-193 & 00:04:08.06 & +14:11:50.71 & 0.58 & 2.90 & 19.81 & 12.55 & 1.48 & 1.26 & 173.64 & 0.57 & C \\
5860-56046-853 & 13:05:52.60 & +20:30:29.08 & 0.63 & 2.64 & 19.82 & 15.77 & 2.51 & 7.81 & 208.01 & 0.77 & C \\
5782-56272-549 & 09:50:59.07 & +20:23:10.61 & 0.65 & 2.21 & 19.77 & 22.94 & 2.28 & 1.56 & 281.30 & 1.29 & C \\
3945-55648-275 & 15:24:18.57 & +18:03:58.60 & 0.60 & 2.26 & 19.44 & 20.38 & 2.34 & 8.23 & 262.28 & 1.19 & C \\
7037-56570-769 & 00:21:23.23 & -04:29:26.17 & 0.56 & 2.15 & 19.77 & 43.13 & 3.07 & 1.22 & 231.63 & 0.95 & $\rm C^{\star}$ \\
6408-56329-327 & 11:18:16.50 & +26:26:48.58 & 0.45 & 2.73 & 19.15 & 8.72 & 1.96 & 8.95 & 175.24 & 0.63 & C \\
6629-56365-311 & 13:55:59.42 & +44:42:43.37 & 0.51 & 2.91 & 19.74 & 6.58 & 1.91 & 2.58 & 186.38 & 0.69 & C \\
6460-56334-699 & 10:07:53.34 & +34:14:19.42 & 0.54 & 2.90 & 19.06 & 10.30 & 2.09 & 6.19 & 196.88 & 0.76 & C \\
4446-55589-635 & 08:19:36.04 & +32:16:45.41 & 0.70 & 2.80 & 19.92 & 10.35 & 2.54 & 5.28 & 172.29 & 0.50 & C \\
6111-56270-647 & 00:10:16.45 & +17:58:52.16 & 0.56 & 2.28 & 19.63 & 11.77 & 1.67 & 8.75 & 261.03 & 1.23 & C \\
5461-56018-391 & 14:21:40.44 & +14:37:49.00 & 0.52 & 2.71 & 19.72 & 12.42 & 3.44 & 7.45 & 254.85 & 1.27 & C \\
4208-55476-395 & 23:04:34.50 & -00:10:58.19 & 0.56 & 2.82 & 19.78 & 8.56 & 1.70 & 2.24 & 184.78 & 0.65 & C \\
5403-55979-51 & 12:35:36.97 & +10:49:23.02 & 0.62 & 2.28 & 19.41 & 28.33 & 3.37 & 5.09 & 366.06 & 2.28 & C \\
6316-56483-84 & 16:19:28.74 & +49:20:55.93 & 0.29 & 2.76 & 17.95 & 17.19 & 2.89 & 4.23 & 242.92 & 1.38 & C \\
4397-55921-505 & 02:00:25.74 & -07:56:31.93 & 0.26 & 2.49 & 17.78 & 16.44 & 2.50 & 5.23 & 238.05 & 1.34 & C \\
5648-55923-862 & 00:13:21.42 & +11:40:39.61 & 0.35 & 2.18 & 18.34 & 20.36 & 2.00 & 5.62 & 223.09 & 1.07 & C \\
\hline
\end{tabular}
%---------------------------------------------------------------------------------
\end{table*}
%----------------------------------------------double peak finished

\section{DATA AVAILABILITY}
The data that support the findings of this study are available on request from the corresponding author. Most of the data is already openly available on the SDSS DR16 website, \url{http://skyserver.sdss.org/dr16/en/tools/explore/Summary.aspx?};

\section*{Acknowledgements}

We thank for Yun Chen and Zhengya Zheng for their helpful suggestions. 
We acknowledge the support from National Key Program for Science and Technology Research and Development of China (2017YFB0203300), National Natural Science Foundation of China (Nos. 11773032, 118513, 11821303, 12022306, 11761131004 and 11761141012),the National Key Research and Development Program of China (No. 2018YFA0404501), RL is supported by NAOC Nebula Talents Program. 

Funding for the Sloan Digital Sky Survey IV has been provided by the Alfred P. Sloan Foundation, the U.S. Department of Energy Office of Science, and the Participating Institutions. SDSS-IV acknowledges
support and resources from the Center for High-Performance Computing at
the University of Utah. The SDSS web site is www.sdss.org.

SDSS-IV is managed by the Astrophysical Research Consortium for the 
Participating Institutions of the SDSS Collaboration including the 
Brazilian Participation Group, the Carnegie Institution for Science, 
Carnegie Mellon University, the Chilean Participation Group, the French Participation Group, Harvard-Smithsonian Center for Astrophysics, 
Instituto de Astrof\'isica de Canarias, The Johns Hopkins University, Kavli Institute for the Physics and Mathematics of the Universe (IPMU) / 
University of Tokyo, the Korean Participation Group, Lawrence Berkeley National Laboratory, 
Leibniz Institut f\"ur Astrophysik Potsdam (AIP),  
Max-Planck-Institut f\"ur Astronomie (MPIA Heidelberg), 
Max-Planck-Institut f\"ur Astrophysik (MPA Garching), 
Max-Planck-Institut f\"ur Extraterrestrische Physik (MPE), 
National Astronomical Observatories of China, New Mexico State University, 
New York University, University of Notre Dame, 
Observat\'ario Nacional / MCTI, The Ohio State University, 
Pennsylvania State University, Shanghai Astronomical Observatory, 
United Kingdom Participation Group,
Universidad Nacional Aut\'onoma de M\'exico, University of Arizona, 
University of Colorado Boulder, University of Oxford, University of Portsmouth, 
University of Utah, University of Virginia, University of Washington, University of Wisconsin, 
Vanderbilt University, and Yale University.

%%%%%%%%%%%%%%%%%%%%%%%%%%%%%%%%%%%%%%%%%%%%%%%%%%



\bibliographystyle{mnras}
\bibliography{lens_search} % if your bibtex file is called example.bib

%\begin{thebibliography}{99}
%\bibitem[\protect\citeauthoryear{Santos, et al.}{2020}]{2020MNRAS.493..141S} Santos S., et %al., 2020, MNRAS, 493, 141
%\bibitem[\protect\citeauthoryear{Bian \& Fan}{2020}]{2020MNRAS.493L..65B} Bian F., Fan X., %2020, MNRAS, 493, L65
%\end{thebibliography}

%%%%%%%%%%%%%%%%%%%%%%%%%%%%%%%%%%%%%%%%%%%%%%%%%%

%%%%%%%%%%%%%%%%% APPENDICES %%%%%%%%%%%%%%%%%%%%%

%\appendix

%%%%%%%%%%%%%%%%%%%%%%%%%%%%%%%%%%%%%%%%%%%%%%%%%%

% Don't change these lines
\bsp	% typesetting comment
\label{lastpage}
\end{document}